\documentclass[12pt,fleqn,titlepage]{article}
\usepackage{geometry}
\usepackage{enumerate}
\usepackage[round]{natbib}
\usepackage{graphicx, graphics}
\usepackage{amssymb}
\usepackage{epstopdf}
\usepackage{amsmath}
\usepackage{tensor}
\usepackage{url}

\begin{document}
\title{Age-Specific Mortality and Fertility Rates for Probabilistic
Population Projections}
\author{Hana \v{S}ev\v{c}\'{\i}kov\'{a}, University of Washington \\
   Nan Li, United Nations\\
Vladim\'{\i}ra Kantorov\'{a}, United Nations \\
Patrick Gerland, United Nations \\
Adrian E. Raftery, University of Washington
  }
\maketitle

\begin{abstract}
The UN released official probabilistic population projections (PPP) for all 
countries for the first time in July 2014. These were obtained by projecting
the period total fertility rate (TFR) and life expectancy at birth ($e_0$)
using Bayesian hierarchical models, yielding a large set of future 
trajectories of TFR and $e_0$ for all countries and future time periods
to 2100, sampled from their joint predictive distribution. 
Each trajectory was then converted to age-specific mortality
and fertility rates, and population was projected using the cohort-component
method. This yielded a large set of trajectories of future age- and
sex-specific population counts and vital rates for all countries.
In this paper we describe the methodology used for deriving the
age-specific mortality and fertility rates in the 2014 PPP, 
we identify limitations of these methods, and we propose several
methodological improvements to overcome them.
The methods presented in this paper are implemented in the publicly
available {\tt bayesPop} R package. \\

\end{abstract}

\begin{small}
\newpage
\tableofcontents

\newpage

\listoffigures
\end{small}

\newpage

\baselineskip=18pt

\section{Introduction}

The United Nations released official probabilistic population projections
for all countries for the first time in July 2014
\citep{gerland&2014}. They were produced by probabilistically
projecting the period total fertility rates (TFR)  and life expectancies 
($e_0$) for all countries using Bayesian hierarchical models 
\citep{Alkema&2011, RafteryChunn&2013}.
These probabilistic projections took the form of a large set of
trajectories, each of which was sampled from the joint predictive
distribution of TFR and female and male $e_0$ for all countries
and all future time periods to 2100 using Markov chain Monte Carlo (MCMC)
methods \footnote{This general approach applies to countries experiencing normal mortality trends. For countries having ever experienced 2 per cent or more adult HIV prevalence during the period 1980 to 2010, all projected trajectories of life expectancy by sex for each of these countries were adjusted in such a way as to ensure that the median trajectory for each country was consistent with the 2012 Revision of the World Population Prospects deterministic projection that incorporates the impact of HIV/AIDS on mortality, as well as assumptions about future potential improvements both in the reduction of the epidemic and survival due to treatment}.

For each trajectory, the life expectancies were converted to 
age- and sex-specific mortality rates, and the total fertility rates
were converted to age-specific fertility rates.
The population was then projected forward using the cohort-component method.
This yielded a large set of trajectories of population by age and sex,
and age-specific fertility and mortality rates, for all countries and
future time periods jointly. These were summarized by 
predictive medians and 80\% and 95\% prediction intervals for 
a wide range of population quantities of interest, for all countries
and a wide range of regional and other aggregates.
They were published as the UN's Probabilistic Population Projections (PPP),
and are available at
\url{http://esa.un.org/unpd/ppp}.

This paper focuses on the methods used to convert probabilistic projections
of $e_0$ and TFR to probabilistic projections of age-specific mortality
and fertility rates. Some limitations of the methods used for the 2014
PPP are identified, and several improvements are proposed to overcome them.
The methods presented in this paper are implemented in an open source R package called bayesPop~\citep{bayespop, Sevcikova2014}.

The paper is organized as follows. In Section \ref{sect-mort} we
describe the current method in PPP for projecting age-specific mortality rates
and our proposed improvements.
In Section \ref{sect-mort.LM} we
outline the Probabilistic Lee-Carter method used in the 2014 PPP.
In the rest of Section \ref{sect-mort}
we propose several improvements to overcome limitations of this method.
These include a new Coherent Kannisto Method for joint projection of 
future age-specific mortality rates at very high ages that avoids 
unrealistic crossovers between the sexes (Section \ref{sec:coherent-kannisto}),
application of the Coherent Lee-Carter method to avoid crossovers
at lower ages (Section \ref{sect-mort.coherentLC}), 
new methods for avoiding jump-off bias (Section \ref{sect-mort.jump-off}),
and application of the Rotated Lee-Carter method to reflect the
fact that when mortality rates are low, they tend to decline faster
at older than at younger ages (Section \ref{sect-mort.rotated}).
In Section \ref{sect-fert}, we describe the current method in PPP for
projecting age-specific fertility rates and our proposed improvements.
We conclude with a discussion in Section \ref{sect-discussion}.

\section{Age-Specific Mortality Rates for Probabilistic Population Projections}
\label{sect-mort}
\subsection{Probabilistic Lee-Carter Method}
\label{sect-mort.LM}
Our methodology is based on the Lee-Carter model \citep{LeeCarter1992}:
\begin{equation*}
\log[m_x(t)] = a_x + b_x k(t) + \varepsilon_x(t), \quad \varepsilon_x(t) \sim N(0, \sigma_{\varepsilon}^2) ,
\end{equation*}
where $m_x(t)$ is the mortality rate for age $x$ and time period $t$. 
The quantity $a_x$ represents the baseline pattern of mortality by age 
over time, and $b_x$ is the average rate of change in mortality rate by 
age group for a unit change in the mortality index $k(t)$.
The parameter $k(t)$ is a time-varying index of the overall level of mortality,
and $\varepsilon_x(t)$ is the residual at age $x$ and time $t$. 
Throughout this paper, $\log$ denotes the natural logarithm.

For a given matrix of rates $m_x(t)$ the model is estimated by a least squares method. The baseline mortality pattern $a_x$ is estimated as the
average of $\log[m_x(t)]$ over the past time periods with observed data.
Since the model is underdetermined, $b_x$ is identified by setting 
$\sum_x b_x = 1$, where the sum is over all ages or age groups $x$.
Also, $k(t)$ is identified by setting  $\sum_{t=1}^T k(t) = 0$, 
where $T$ is the number of past time periods for which data are available. 
The estimates are then
\begin{eqnarray}
\hat{a}_x &  = & \frac{\sum_{t=1}^T \log[m_x(t)]}{T} \label{eq:ax}\\
\hat{k}(t)  & = & \sum_x \left\{\log[m_x(t)] - \hat{a}_x\right\} \label{eq:kt}\\
\hat{b}_x  & = & \frac{\sum_{t=1}^T \left\{\log[m_x(t)] - \hat{a}_x\right\}
\hat{k}(t) }{\sum_{t=1}^T \hat{k}(t)^2} . \label{eq:bx}
\end{eqnarray}

To forecast $m_x(t)$, one needs to project $k(t)$ into the future.
To project $k(t)$, the Lee-Carter method uses a random walk with drift:
\begin{equation*}
\hat{k}(t+1) = \hat{k}(t) + \hat{d}, \quad \text{where} \; 
  \hat{d}=\frac{1}{T-1}[\hat{k}(T) - \hat{k}(1)] .
\end{equation*}
 \citet{LeeMiller2001} proposed replacing the step of projecting $k(t)$ by itself by matching future $k(t)$ to future projected $e_0(t)$. 

Current calculations are done using a highest age or open interval of 85$+$. 
For projections one needs to 
 extend mortality rates to higher ages $x$, usually beyond $100+$, 
because mortality rates are expected broadly to decline over time in the 
future, so there will be larger numbers of people at higher ages. 
For extending the force of mortality at older age groups, the Kannisto model provides a robust way to fit available mortality rates from age 80 to 100, and to extrapolate mortality rates up to age 130 in a way that is consistent with empirical observations on oldest-old mortality \citep{Thatcher1998}.
 
The Bayesian probabilistic projections of life expectancy 
\citep{RafteryChunn&2013,RafteryLalic2014} provide us with 
a set of  future trajectories of female and male $e_0$, 
representing a sample from the joint predictive
distribution of future female and male $e_0$ for all countries and all
future time periods. 
The 2014 PPP used methods for turning 
a trajectory of future $e_0$ values into a set of future age-specific 
mortality based on the ideas of \citet{LeeMiller2001} 
and \citet{LiGerland2011}; see \citet{Raftery&2012}.
They were based on the following algorithm:

\subsubsection*{Algorithm 1}
Let $t \in \{1,\ldots,T\}$ and $\tau \in \{T+1, \dots, T_p \}$ denote the observed and projected time periods, respectively. 
\begin{enumerate}
\item Using the Kannisto method extend $m_x(t)$ to higher age groups so that $\max(x) = 130+$ for all $t$.
\item Estimate $a_x$, $k(t)$ and $b_x$ using the extended historical $m_x(t)$ 
(equations \ref{eq:ax}--\ref{eq:bx}).
\item For a given $e_0(\tau)$ in each trajectory and given $a_x$ and $b_x$, 
solve for future $k(\tau)$ numerically using life tables. 
This yields a nonlinear equation which can be solved using 
the bisection method. More details are given in Section~\ref{sec:lifetables}.
\item Compute mortality rates by 
$\log[m_x(\tau)] = \hat{a}_x + \hat{b}_x k(\tau)$ 
for each trajectory and future time $\tau$.
\end{enumerate}
Applying these steps to all trajectories of $e_0$ yields a posterior 
predictive distribution of $m_x(t)$.

However, this procedure has a number of drawbacks.
There is no assurance that the extension of $m_x(t)$ to higher ages 
yields mortality rates that are coherent between males and females. 
Similarly, the predicted $m_x(\tau)$ can lead to unwanted crossovers
between female and male mortality rates,
since they are obtained independently for each sex.
In the following sections, we present solutions to these and other 
limitations of the simple algorithm above, and give more details about Step 3.

\subsection{Coherent Kannisto Method}
\label{sec:coherent-kannisto}
A sex-independent extension of the observed mortality rates 
to higher age categories can lead to unrealistic crossovers at higher ages.
We propose a modification of the Kannisto method that treats male and female mortality rates jointly. In this section, for simplicity we omit the time index $t$.

The original Kannisto model has the form
\begin{eqnarray*}
m_x & = & \frac{c e^{dx}}{1+c e^{dx}} e^{\varepsilon_x}, \; \mbox{ or } \\
\text{logit}(m_x) & = & \log c + dx + \varepsilon_x ,  
\end{eqnarray*}
where $\varepsilon_x$ is a random perturbation with mean zero.
The model is usually estimated independently for each sex, assuming independence
across ages and normality of 
the $\varepsilon_x$, using a maximum likelihood method \citep{Thatcher1998, Wilmoth2007}. This yields sex-specific parameter estimates $\hat{d}_M$, $\hat{d}_F$, $\hat{c}_M$, $\hat{c}_F$.

We suggest modifying this by forcing the sex-specific parameters
$d_M$ and $d_F$ to be equal (i.e. $d_M=d_F=d$), but still allowing the 
parameters $c_M$ and $c_F$ to differ between the sexes:
\begin{equation*}
\text{logit}(m_x^g)  =  \log c_g + dx + \varepsilon_{x}^g, 
 \quad \mbox{ for } \; g=M, F.
\end{equation*}
This leads to the following model:
\begin{equation*}
\text{logit}(m_x^g) = \beta_0 + \beta_1 1_{(g=M)} + \beta_2 x + \varepsilon_{x}^g , 
\end{equation*}
where $1_{(g=M)} = 1$ if $g=M$ and 0 otherwise.

To estimate the $\beta$ parameters, we fit the model to the observed $m_x$ 
for ages 80--99 by ordinary least-squares regression, which corresponds to 
maximum likelihood under the assumptions of independence and normality of
the $\varepsilon_x^g$. There are four age groups in the data used for
fitting the model, and thus eight points in total for both sexes. Then,
\begin{eqnarray*}
\hat{c}_F & = & e^{\hat{\beta}_0} , \\
\hat{c}_M & = & e^{\hat{\beta}_0 + \hat{\beta}_1} , \\
\hat{d} & = & \hat{\beta}_2 . 
\end{eqnarray*}

\begin{figure}[ht]
\begin{center}
\includegraphics[width=1.0\textwidth]{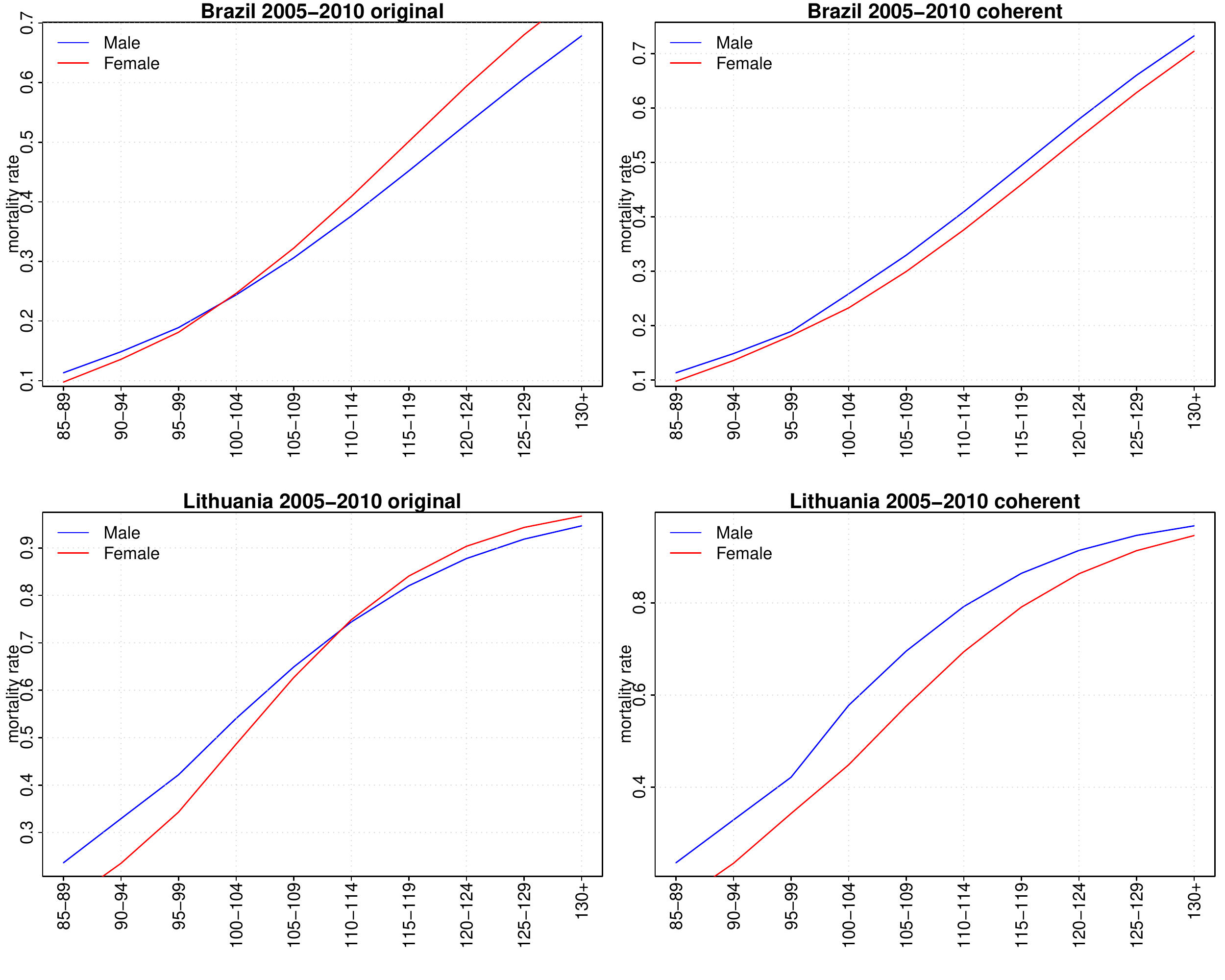}
\caption[Original and Coherent Kannisto methods compared:
Male and female mortality rates for Brazil and Lithuania]
{\label{fig:kannisto}Mortality rates for male  (blue) and female (red) extended using the coherent Kannisto method (right panels) compared to original Kannisto (left panels) for Brazil and Lithuania in the last observed time period (2005-2010). }
\end{center}
\end{figure}

Figure~\ref{fig:kannisto} shows the resulting $m_x$ for old ages for Brazil 
and Lithuania in the last observed time period. 
From the left panels we see that there are crossovers using the 
classic Kannisto method, which is unrealistic.
However, male mortality stays above female mortality in the coherent
version, as can be seen in the right panels; this is more realistic.

\subsection{Coherent Lee-Carter Method}
\label{sect-mort.coherentLC}
We will adopt an extension of the Lee-Carter method suggested by \citet{Li2005}, the so-called {\em coherent} Lee-Carter method. It takes into account the fact that mortality patterns for closely related populations are expected to be similar. In our application, these related populations will be males and females in the same country, since there is no expectation that the life expectancy will diverge between such groups. Thus, the Lee-Carter method is extended by two requirements:
\begin{eqnarray}
b_x^M & = & b_x^F , \nonumber \\
k_M(\tau) & = & k_F(\tau) , \label{eq:coherent-k}
\end{eqnarray}
where $M$ and $F$ denotes male and female sex, respectively. This ensures that the rates of change of the future mortality rates are the same for the two 
sexes, and thus avoids crossovers.

\subsection{Avoiding Jump-off Bias}
\label{sect-mort.jump-off}
Mortality rates in the last period of the historical data used
for estimation (or jump-off period) are 
commonly referred to as jump-off rates 
\citep{Booth2006}. Often there is a mismatch between fitted rates for the 
last period $T$ and the actual rates (jump-off bias). As a result, a discontinuity between the actual rates in the jump-off period and the rates projected in the first projection period may occur. 

A possible solution to avoid jump-off bias is to constrain the model in 
such a way that $k(t)$ passes through zero in the jump-off period $T$,
and to use $m_x$  only from the last fitting period to obtain 
$a_x$ \citep{LeeMiller2001}:
\begin{eqnarray}
a_x & = & \log[m_x(T)] \implies k(T) = 0 . \label{eq:ax-latest}
\end{eqnarray}

 A disadvantage of this solution is that in cases where the mortality 
rates are bumpy in the jump-off period (i.e. not smooth across ages),
this ``bumpiness" propagates into the future.  In general for projections, 
we suggest using the age-specific mortality rates from the 
last fitting period and smoothing them over age if necessary 
(e.g. for small populations with few deaths in some age groups) 
while preserving the value for the youngest age group:
\begin{eqnarray}
a_x & = & \text{smooth}_x\{\log[m_x(T)] \} \; \text{ with } \; 
 a_{0-1} = \log[m_{0-1}(T)] .  \label{eq:ax-latest-smooth}
\end{eqnarray}

\begin{figure}[ht]
\begin{center}
\includegraphics[width=\textwidth]{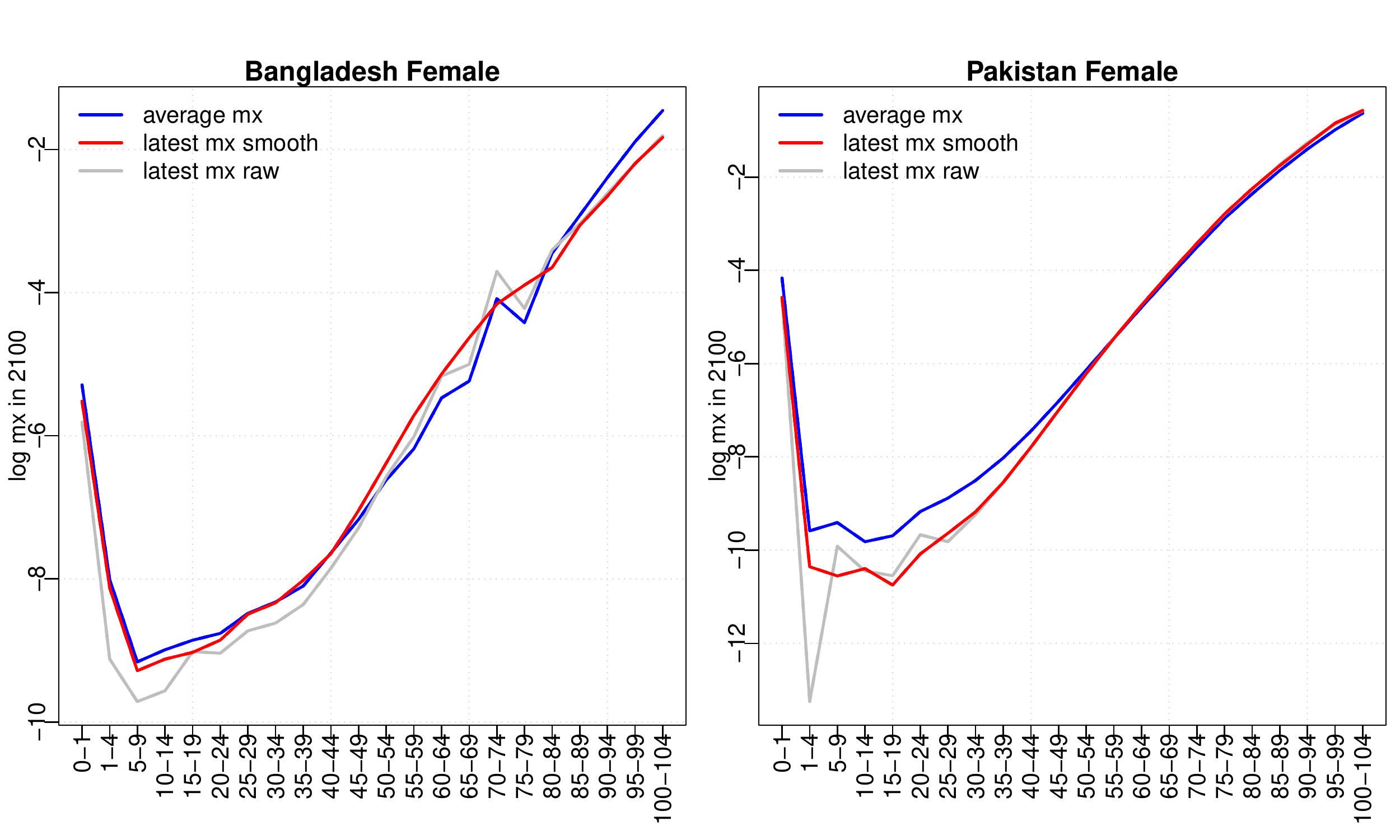}
\caption[Projected female mortality rates for Bangladesh and Pakistan in 
2095-2100 projected using three different methods for computing $a_x$]
{\label{fig:jumpoff}Log female mortality rates for Bangladesh and 
Pakistan in 2095-2100 projected using three different methods for computing 
$a_x$: (1) using an average $m_x$ over time (blue line); 
(2) using the latest smoothed $m_x$ (red line); and
(3) using the latest $m_x$ as it is (grey line).}
\end{center}
\end{figure}

Figure~\ref{fig:jumpoff} shows the resulting difference in $\log[m_x(\tau)]$ projected to $\tau=2095-2100$ for two countries using the three different methods of computing $a_x$, namely equations (\ref{eq:ax}), (\ref{eq:ax-latest}) and 
(\ref{eq:ax-latest-smooth}). As can be seen in the case of Bangladesh, 
the smoothing step removes bumps whereas the averaging method does not.

\begin{figure}[ht]
\begin{center}
\includegraphics[width=\textwidth]{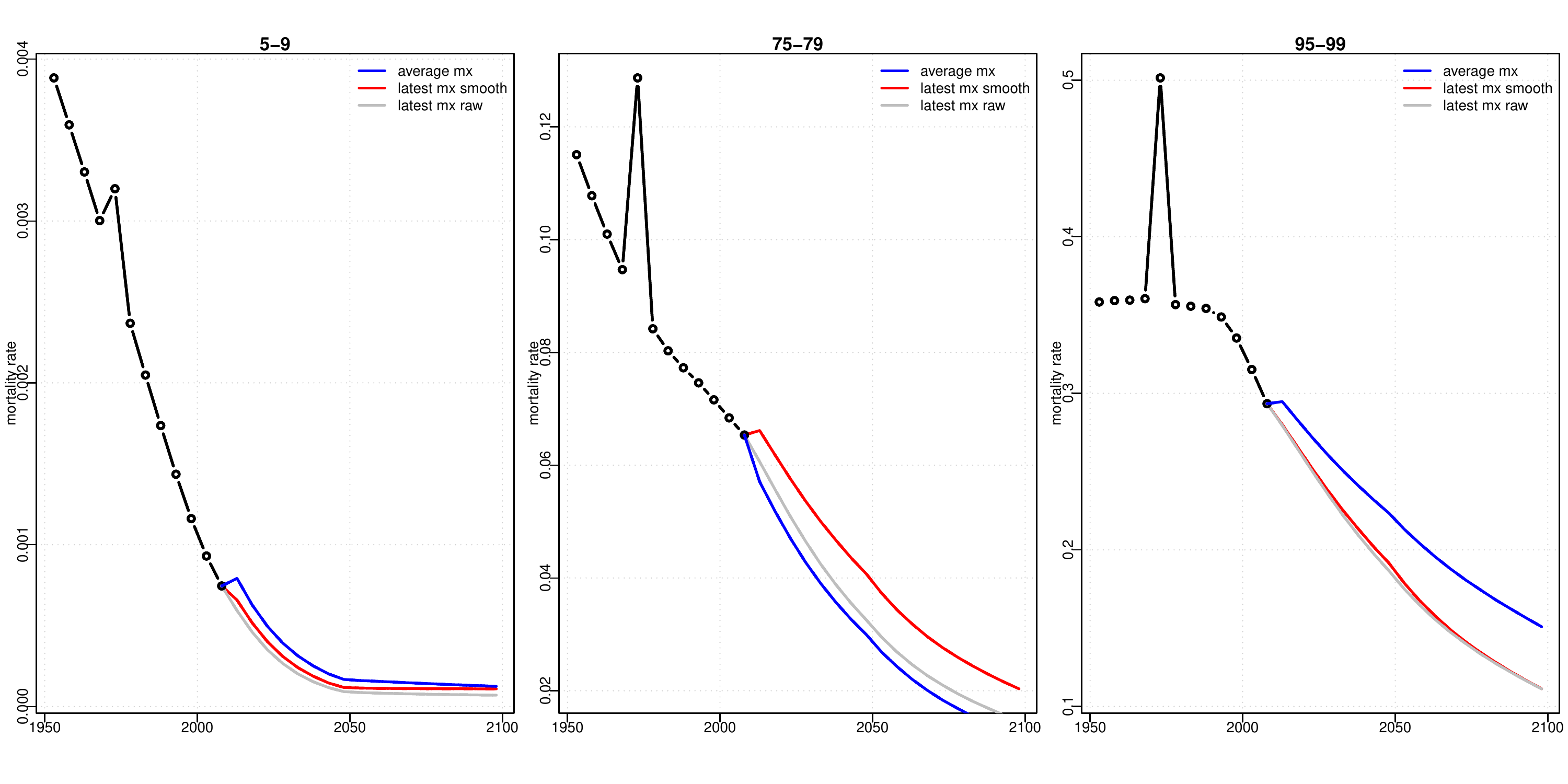}
\end{center}
\caption[Mortality rates  for Bangladesh by time for three different age groups]
{\label{fig:jumpoff-time}Mortality rates  for Bangladesh by time for three different age groups. Colors correspond to the same methods as in Figure~\ref{fig:jumpoff}.}
\end{figure}

Figure~\ref{fig:jumpoff-time} shows the impact of the methods on $m_x$ as time series for Bangladesh for three different age groups. Using the average $m_x$ results in  jump-offs for the 5--9 and 95--99 age groups (blue curve). If the latest raw  $m_x$ are used, the jump-offs are eliminated (grey curve). 
A smoothed version creates a new jump-off for the age group 75--79 (red curve).

This shows that there is a trade-off between bumpy mortality rates over ages in later projection years and no jump-offs, and smooth mortality rates with no jump-offs. Our solution is to decide  on  a country-specific basis which method is more appropriate.

\subsection{Rotated Lee-Carter Method}
\label{sect-mort.rotated}
\citet{LiLeeGerland2013} focused on the fact that in more developed regions, once countries have already reached a high level of life expectancy at birth, the mortality decline decelerates at younger ages and accelerates at old ages. 
This change in the pace of mortality decline by age cannot be captured by 
the original Lee-Carter method, since this constrains the rate of
change $b_x$ to be constant over time. They proposed instead
rotating the $b_x$ over time to a so-called {\em ultimate} $b_x$, 
denoted by $b_{u,x}$, which is computed as follows.

Let 
\[
\bar{b}_{15-64} = \frac{1}{10}\sum_{x=15-19}^{60-64}b_x \, . 
\]
Then 
\begin{eqnarray}
b_{u,x} & = & \left\{
\begin{array}{ll}
\bar{b}_{15-64} & \text{ for } \; x\in\{0-1, 1-4, 5-9, \dots, 60-64\}, \\
b_x \cdot b_{u,60-64} /  b_{65-70} & 
  \text{ for } \; x\in\{65-70, \dots, 130+ \} ,
\end{array}\right.
\label{eq:bux}
\end{eqnarray}
with $b_{u,x}$ scaled to sum to unity over all ages.

The rotation is dependent on $e_0(\tau)$, and so the resulting $b_x$ also 
becomes time-dependent. The rotation finishes at a certain level of live expectancy, denoted by $e_0^u$. \citet{LiLeeGerland2013}  
recommend using $e_0^u = 102$. Using the smooth weight function
\begin{eqnarray*}
w(\tau) & = & \left\{ \frac{1}{2}\left[1+\sin[\frac{\pi}{2}(2w'(\tau)-1)]\right]\right\}^\frac{1}{2} \quad \text{with} \quad
w'(\tau)  =  \frac{e_0(\tau) - 80}{e_0^u - 80} ,
\end{eqnarray*}
the rotated $b_x$ at time $\tau$, denoted by $B_x(\tau)$, is derived as:
\begin{eqnarray}
B_x(\tau) & = & \left\{ 
\begin{array}{lc}
b_x, & e_0(\tau) < 80, \\
\left[1-w(\tau)\right]b_x + w(\tau)b_{u,x}, & 80 \leq e_0(\tau) < e_0^u , \\
b_{u,x}, & e_0(\tau) \geq e_0^u .
\end{array}\right.
\label{eq:Bxt}
\end{eqnarray}

\begin{figure}[ht]
\begin{center}
\includegraphics[width=\textwidth]{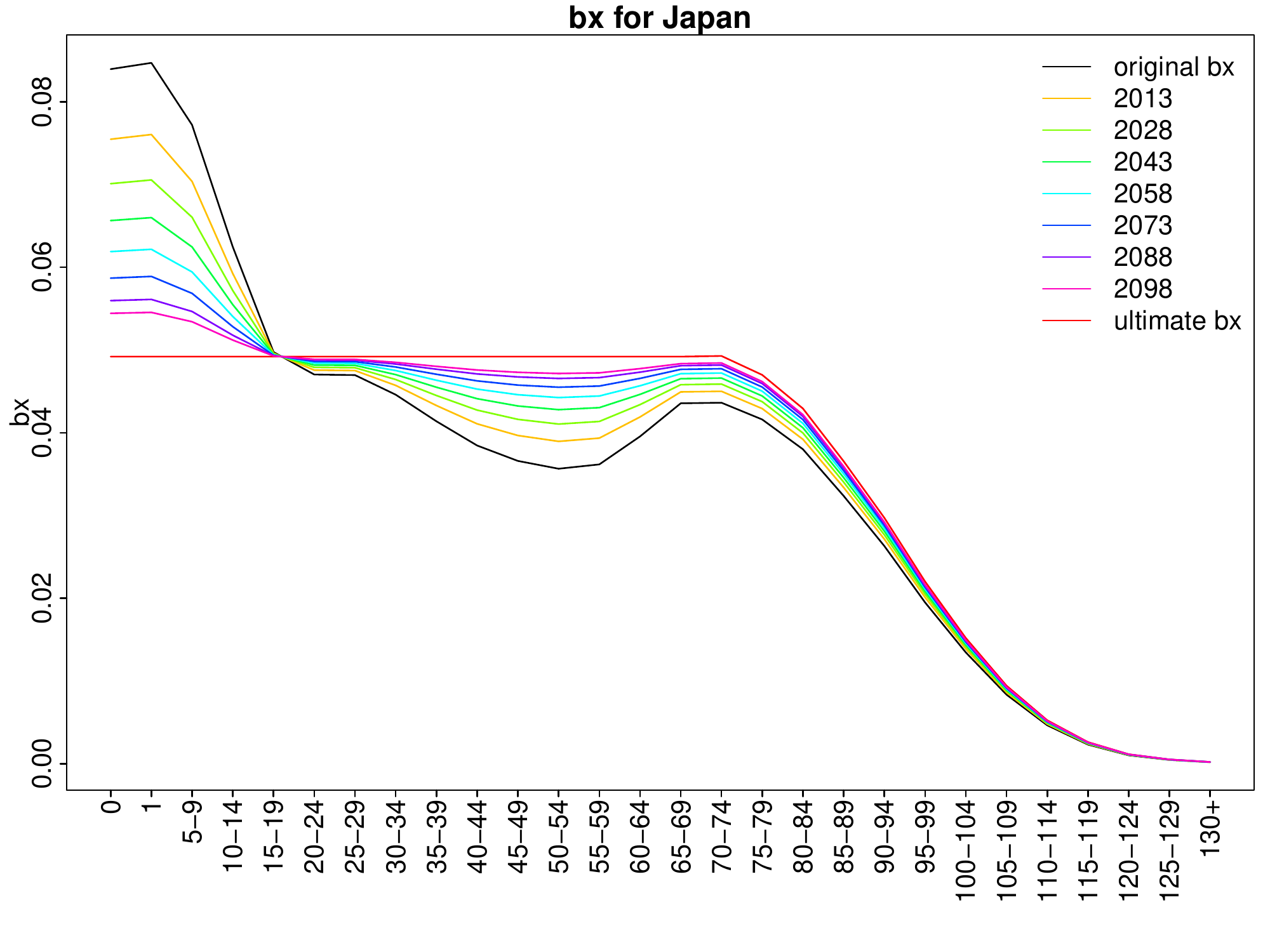}
\caption[Rotating the parameter $b_x$ over time: Example from Japan]
{\label{fig:rotation-Japan}Rotating the parameter $b_x$ over time. The original $b_x$ (black curve) approaches the ultimate $b_{u,x}$ (red curve) over time starting with yellow and continuing towards the darker colors.}
\end{center}
\end{figure}

Figure~\ref{fig:rotation-Japan} shows the results for Japan as an example. 
The original $b_x$ is shown by the black curve. The ultimate $b_{u,x}$,
to be reached at life expectancy of 102, is in red. The remaining curves show the change over time starting with yellow and continuing towards the red curve. 

\subsection{Computing Life Tables}
\label{sec:lifetables}
Step 3 in Algorithm 1 calls for matching future  $k(\tau)$ to projected $e_0(\tau)$. This is a nonlinear equation in $k(\tau)$.  It is solved by an iterative nonlinear procedure in which for given values of $a_x$, $b_x$ and $k(\tau)$ a life table is produced, and the resulting life expectancy is computed and compared with the projected  $e_0(\tau)$. We used a bisection method to solve the
nonlinear equation. This is simple and robust and involves relatively
few iterations. It would be possible to use a nonlinear solution method
that is more efficient computationally, but the computational gains 
would be modest and this could make the method much more complex.

In the process of computing life tables, 
 the conversion of mortality rates $m_x(\tau)$ to probabilities of dying $q_x(\tau)$ follows the approach used by the United Nations to compute abridged life tables. 
This is computed by the LIFTB function in Mortpak \citep{MortPak1988, MortPak2013}, where at a given time point the probability of dying for an individual between age $x$ and $x+n$ is:
\begin{eqnarray}
_n{q_x} & = & \frac{n * \tensor[_n]{m}{_x}}{1 + (n - \tensor[_n]{A}{_x}) * \tensor[_n]{m}{_x}}, \label{eq:qx}
\end{eqnarray}
with  $n$ being the length of the age interval and $_n{A_x}$ being the average number of years lived between ages $x$ and $x+n$ by those dying in the interval. With  $l_x$ being the number of survivors at age $x$, we have
\begin{eqnarray}
{l_{x+n}} & = & {l_x}{(1 - \tensor[_n]{q}{_x})} ,  \label{eq:lx}\\
_n{d_x} & = & {{l_x} - {l_{x+n}}} ,  \label{eq:dx}\\
_n{L_x} & = & {_n{A_x}{l_x} - {(n - \tensor[_n]{A}{_x})l_{x+n}}} ,  \label{eq:Lx}
\end{eqnarray}
where  $\tensor[_n]{d}{_x}$ denotes the number of deaths between ages $x$ and $x+n$ and  $\tensor[_n]{L}{_x}$ denotes the number of person-years lived between ages $x$ and $x+n$. The expectation of life at age $x$ (in years) $e_x$ 
is given by 
\begin{equation*}
{e_x}  =   \frac{T_x}{{l}_{x}}  \label{eq:Ex} \quad \text{with} \quad {T_x}  =  \sum _{a=x}^{\infty}{_n}{L}{_a} ,
\end{equation*}
 where $T_x$ is the number of person-years lived at age $x$ and older.

For ages 15 and over, the expression for $_n{A_x}$ is derived from the 
\citet{Greville1943} approach  to calculating age-specific separation factors based on the age pattern of the mortality rates themselves with:
\begin{equation*}
_n{A_x} = 2.5 - \frac{25}{12} (_n{m_x} - k), \quad \text{where} \; k = \frac{1}{10} \log \left(\frac{ \tensor[_n]{m}{_{x+5}}}{ \tensor[_n]{m}{_{x-5}}}\right) .
\end{equation*}
For ages 5 and 10, $_n{A_x} = 2.5$ and for ages under 5,  values from the Coale and Demeny West region relationships are used for $_n{A_x}$ \citep{CoaleDemeny1966}.\footnote{The Coale and Demeny West region formulae are used as follows. When $\tensor[_0]{m}{_1} \geqslant 0.107$, then $_1{A_0} = 0.33$ for males and $0.35$ for females; $_4{A_1} = 1.352$ for males and $1.361$ for females. When $_1{m_0} <   0.107$, $_1{A_0} = 0.045 + (2.684 \cdot  \tensor[_1]{m}{_0}$) for males and $_1{A_0} = 0.053 + (2.800 \cdot  \tensor[_1]{m}{_0})$ for females; $_4{A_1} = 1.651 - (2.816 \cdot  \tensor[_1]{m}{_0})$ for males and $_4{A_1} = 1.522 - (1.518 \cdot \tensor[_1]{m}{_0})$ for females.}

\subsection{Summary of Improved Algorithm}
We now summarize the modifications described in the previous sections 
by proposing an improved algorithm for deriving the age-specific mortality 
rates $m_x$ for potential use in future probabilistic population projections.

\subsubsection*{Algorithm 2}
As before, let $t \in \{t_1,\dots, T\}$ and $\tau \in \{T+1, \dots T_p\}$ 
denote the observed and projected time periods, respectively. Also, let $g\in\{F,M\}$ be an index to distinguish sex-specific measures.
\begin{enumerate}
\item Using the Coherent Kannisto Method from Section~\ref{sec:coherent-kannisto}, extend $m_x(t)$ to higher age categories with $\max(x) = 130+$ for all $t$.
\item Choose a method to estimate $a_x$, i.e. one of equations 
(\ref{eq:ax}), (\ref{eq:ax-latest}) or (\ref{eq:ax-latest-smooth}),
depending on country specifics.\footnote{In the bayesPop package this country-specific set of options is controlled through two dummy variables in the {\tt vwBaseYear2012} dataset: (1) whether the most recent estimate of age mortality pattern should be used (LatestAgeMortalityPattern) and (2) whether it should be smoothed (SmoothLatestAgeMortalityPattern). See {\tt vwBaseYear2012} in R.} Do the estimation for each sex $g$, obtaining $a_x^g$.
\item Estimate $k(t)$ and $b_x$ using the extended historical $m_x(t)$ 
and equations (\ref{eq:kt}--\ref{eq:bx}) for $g=M, F$ independently, 
yielding  $b_x^g$.
\item Given $b_x^M$ and $b_x^F$ from Step 3, set $b_x = \frac{b_x^M + b_x^F}{2}$.
\item Compute the ultimate $b_{ux}$ as in equation (\ref{eq:bux}).
\item For a combined $e_0(\tau) = \left[e_0^M(\tau) + e_0^F(\tau)\right]/2$ in each trajectory, compute $B_x(\tau)$ as in equation (\ref{eq:Bxt}).
\item For a given sex-specific $e_0^g(\tau)$ in each trajectory and given $a_x^g$ and $B_x(\tau)$, solve for future $k_g(\tau)$ numerically using life tables. 
This yields a nonlinear equation which is solved using the bisection method, as described in Section~\ref{sec:lifetables}.
\item For each trajectory, time $\tau$ and sex $g$, compute mortality rates by $\log[m_x^g(\tau)] = a_x^g + B_x(\tau) k_g(\tau)$.
\item Since the previous step does not comply with equation 
(\ref{eq:coherent-k}) and thus can lead to crossovers in high ages, an additional constraint is added:\\
If $e_0^M(\tau) < e_0^F(\tau)$ then 
\begin{eqnarray*}
m_x^M(\tau) &= &\max[m_x^M(\tau), m_x^F(\tau)] \text{ for } x \geq 100  .
\end{eqnarray*}
\end{enumerate}

\begin{figure}[ht]
\begin{center}
\includegraphics[width=\textwidth]{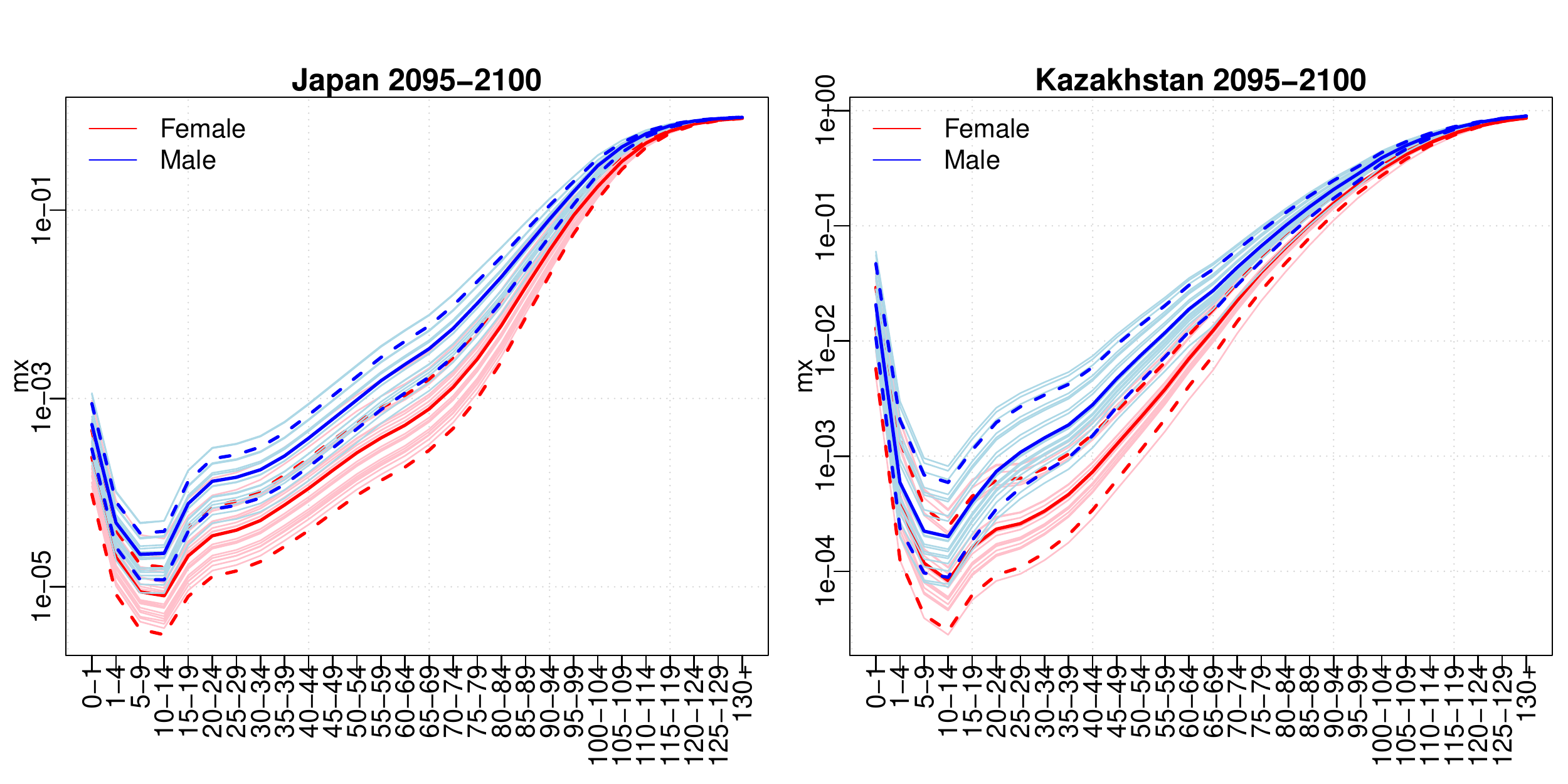}
\caption[Probabilistic projection of age-specific mortality rates for Japan 
and Kazakhstan in 2095-2100]
{\label{fig:prob-mx-2100}Probabilistic projection of age-specific mortality rates for Japan (left panel) and Kazakhstan (right panel) in the time period 2095-2100. Both plots show the marginal distribution for male (blue lines) and female (red lines) where the dashed lines mark the 80\% probability intervals and the solid lines are 20 randomly sampled trajectories (out of 1000) for each sex.  The $y$-axis is on the logarithmic scale.}
\end{center}
\end{figure}

\begin{figure}[ht]
\begin{center}
\includegraphics[width=\textwidth]{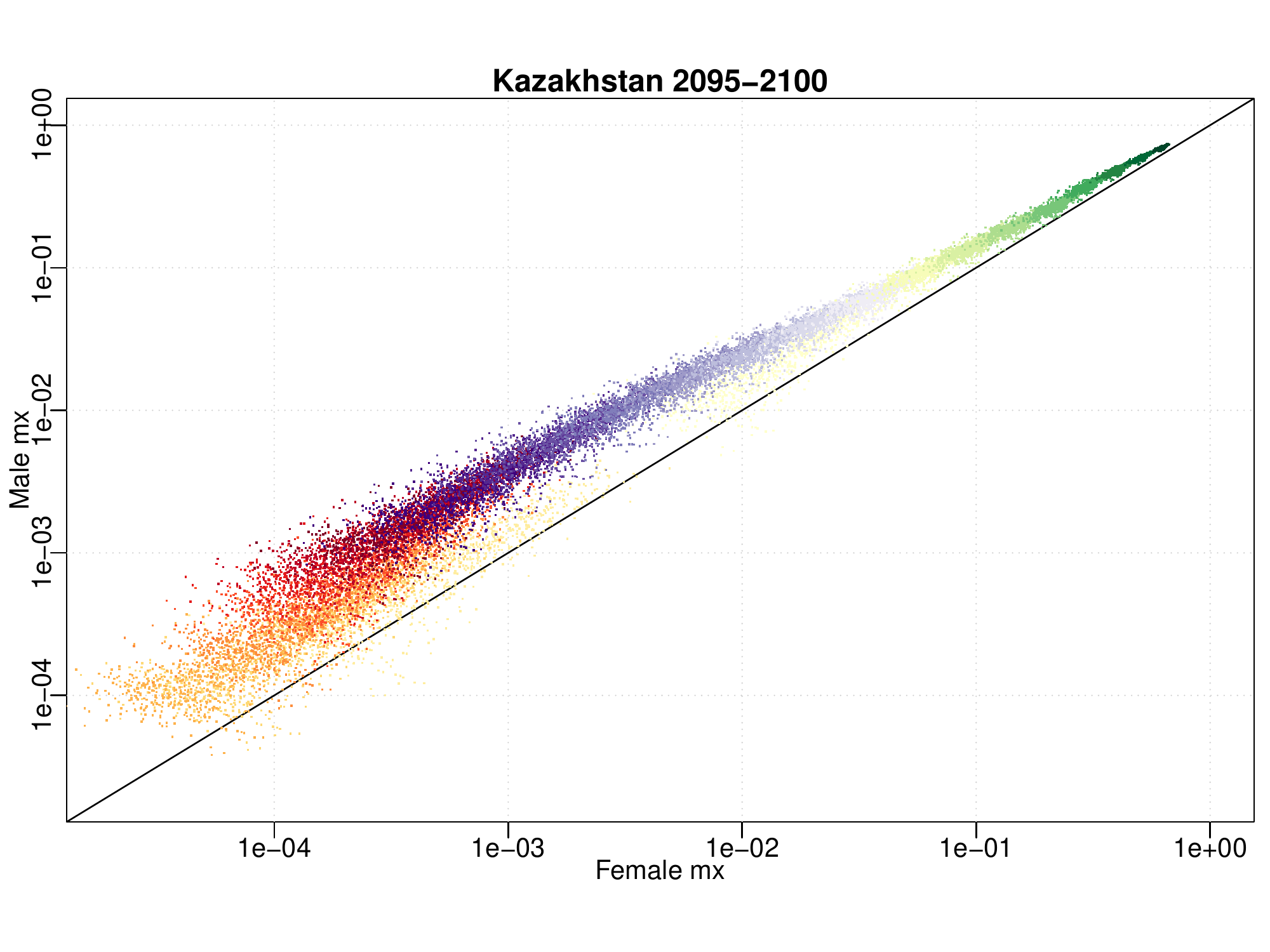}
\caption[Joint predictive distribution of mortality rates for females and males for Kazakhstan in 2095-2100]
{\label{fig:prob-mx-joint-all}The joint predictive distribution of mortality rates for females and males for Kazakhstan in 2095-2100. It shows mortality rates from all age groups where age groups are distinguished by colors. Both axes are on the logarithmic scale. There are 1000 points per age group.}
\end{center}
\end{figure}

\begin{figure}[ht]
\begin{center}
\includegraphics[width=\textwidth]{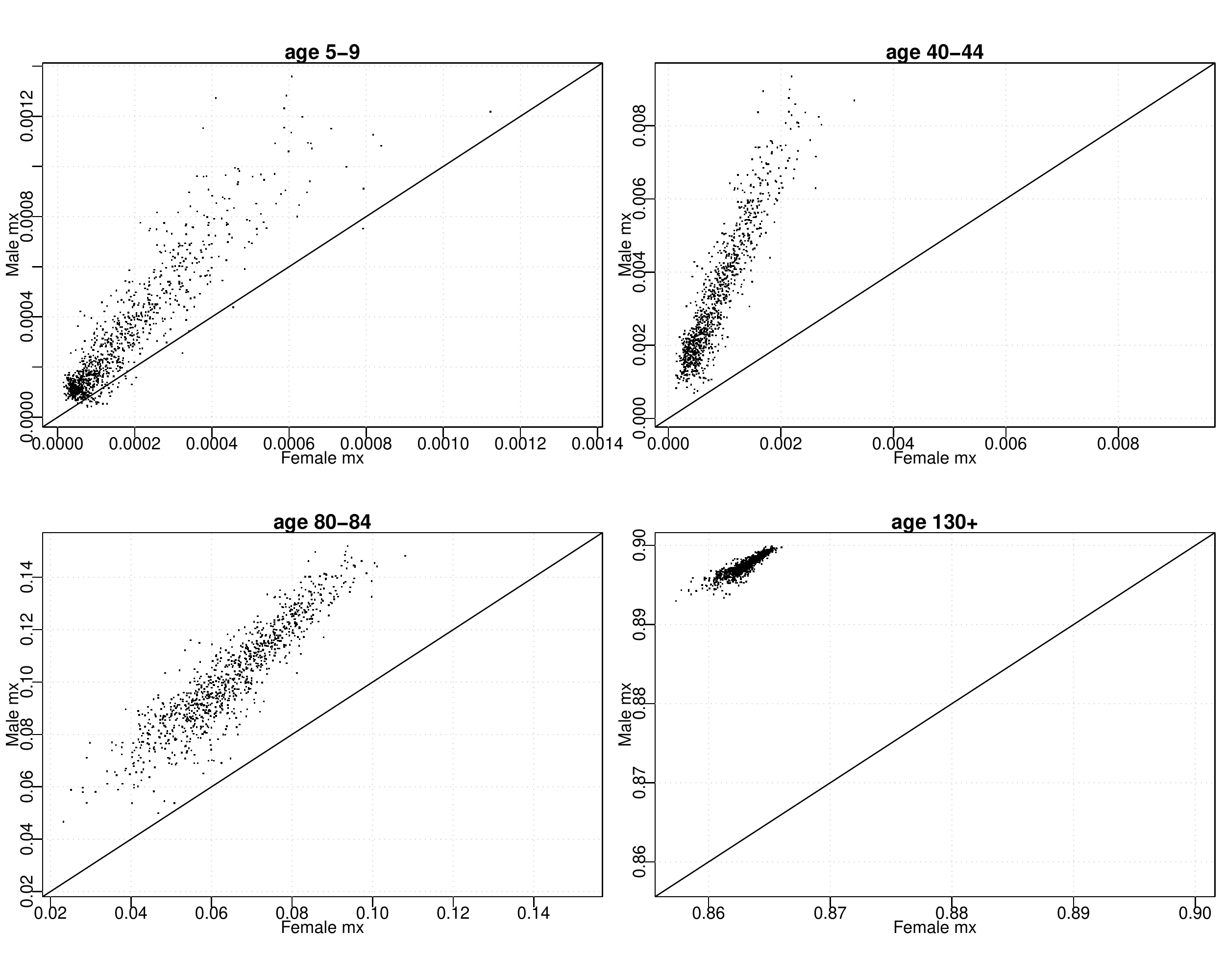}
\caption[Joint distribution of mortality rates for females and males for Kazakhstan in 2095-2100 for individual age groups]
{\label{fig:prob-mx-joint}The joint distribution of mortality rates for females and males for Kazakhstan in 2095-2100 for individual age groups. In all panels, 1000 points are shown and all axes are on a normal scale.}
\end{center}
\end{figure}

Figure~\ref{fig:prob-mx-2100} shows the resulting probabilistic projection of $m_x(\tau)$ for the period 2095-2100 for both sexes in two selected countries. In addition to the marginal distribution for Kazakhstan in the right panel of Figure~\ref{fig:prob-mx-2100}, its joint distribution for males and females 
is shown  in Figure~\ref{fig:prob-mx-joint-all} on a logarithmic scale.
Points below the $x=y$ solid line indicate crossovers in the individual trajectories. It can be seen  that only a few trajectories experience crossovers when mortality is low, i.e. in young ages, suggesting a low (but non-zero) probability for such an event, while there are no crossovers for high mortality, i.e. in old ages. We observed similar results  for most countries. Figure~\ref{fig:prob-mx-joint} shows the same joint distribution for selected age groups on a normal scale.

\subsubsection*{Exceptions}
For about 50 countries, insufficient detailed data about mortality by age and sex are available between 1950 and 2010 \citep{UN2013b}. Therefore, the age patterns of mortality are based on model life tables (e.g., Coale-Demeny). For these countries a model $b_x$ associated with one of the regional model life tables is used (see Table 2 page 18 in \citet{LiGerland2011}).\footnote{In the bayesPop package this country-specific set of options is controlled through two variables in the {\tt vwBaseYear2012} dataset: (1) the type of age mortality pattern used for the estimation period (AgeMortalityType with the option "Model life tables") and (2) the specific mortality pattern used (AgeMortalityPattern with options like "CD West").}

In addition, for about 40 countries with a generalized HIV/AIDS epidemic, age patterns of mortality since the 1980s have been affected by the impact of AIDS mortality (especially before the scaling up of antiretroviral treatment starting in 2005). For these countries the application of the conventional Lee-Carter approach is inappropriate.\footnote{In the BayesPop package this specific-set of countries are identified through a dummy variable (WPPAIDS) in the {\tt vwBaseYear2012} dataset.} Instead, we introduce a modification where steps 2--6 in Algorithm~2 are replaced by the following steps:
\begin{enumerate}
\item Start with the most recent $a_x$ (affected by impact of HIV/AIDS on mortality) and smooth it as in equation (\ref{eq:ax-latest-smooth}), 
obtaining $a_x^s$.
\item Compute an {\em ultimate} (or ``AIDS-free'' target) $a_x$, denoted by $a_x^u$, which is a smoothed average of historical $\log(m_x)$ up to 1985 (i.e., prior to the start of the impact caused by HIV/AIDS on mortality), denoted by $a_x^v$:
\begin{eqnarray}
a_x^v & = & \frac{\sum_{t=t_1}^{t_{T_u}} \log[m_x(t)]}{T_u}  \; \text{ with } \; t_{T_u}=1985 \nonumber \\
a_x^u & = & \text{smooth}_x\{ a_x^v \} \; \text{ with } \; a_{0-1}^u = a_{0-1}^v   \label{eq:ax-aids-smooth}
\end{eqnarray}
\item For each $x$ interpolate from $a_x^s$ to $a_x^u$ assuming that in the long run the excess mortality due to the HIV/AIDS epidemic disappears (or reaches a very low endemic level with negligible mortality impact) both as a result of decreased HIV prevalence, improved access to treatment and survival with treatment.
\item During the projections, pick an $a_x(\tau)$ by moving along the interpolated line of the corresponding $x$, so that $a_x^u$ is reached by 2100.
\item As above, $b_x$ is associated with one of the regional model life tables.
\end{enumerate}

\begin{figure}
\begin{center}
\includegraphics[width=\textwidth]{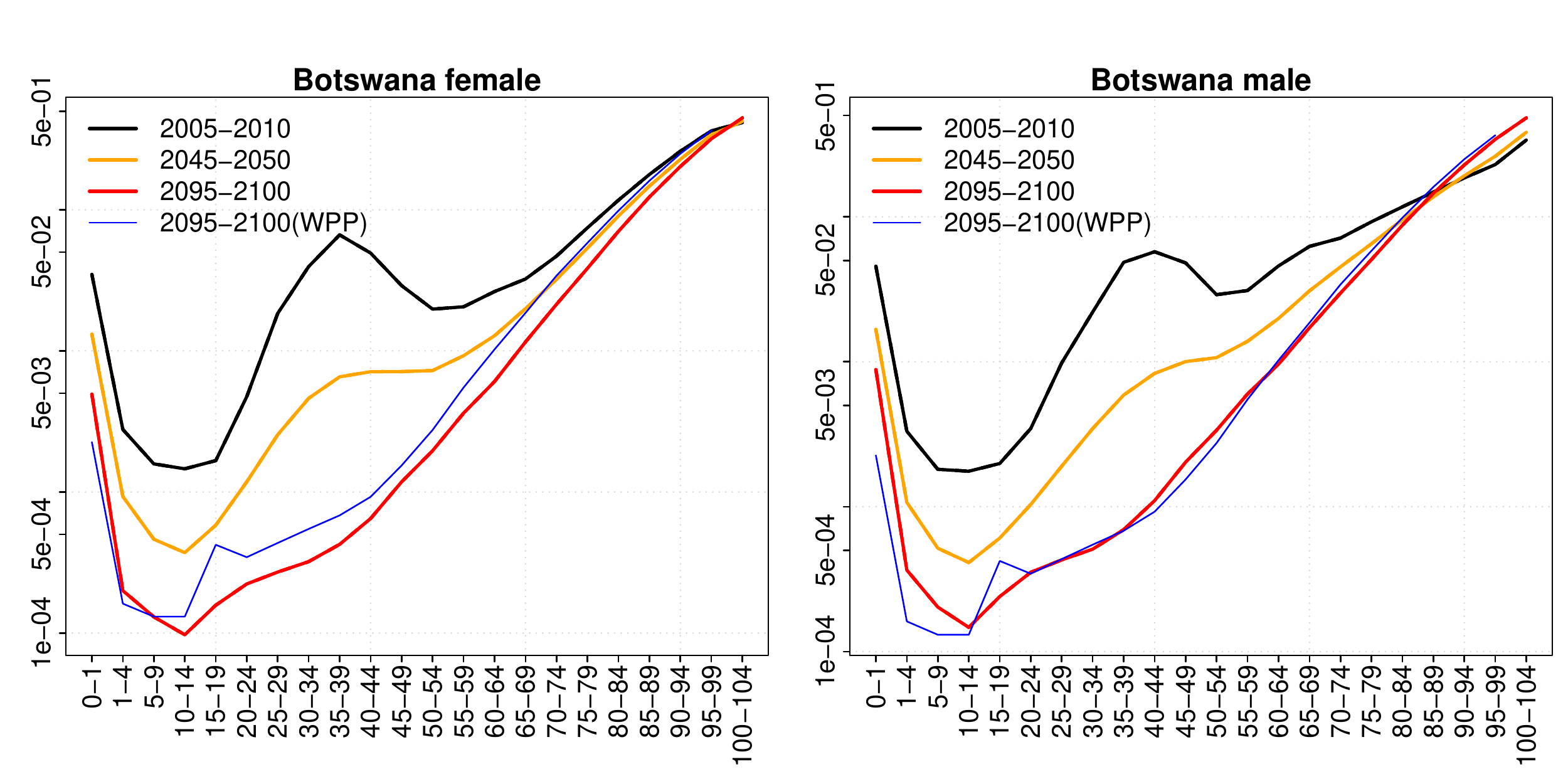}
\end{center}
\caption{\label{fig-Botswana} Projected age-specific mortality rates
for Botswana, a country with a generalized HIV/AIDS epidemic.}
\end{figure}

An example of the resulting projected median age-specific mortality rates
for Botswana, a country with a generalized HIV/AIDS epidemic,
is shown in Figure \ref{fig-Botswana}.

There has been recent progress in the modelling of age patterns of 
mortality for countries with generalized HIV/AIDS \citep{Sharrow2014}. 
This could provide additional options to better incorporate the uncertainty 
about future HIV prevalence, expanded access to treatment, underlying age mortality patterns, and their interaction on overall mortality by age into
probabilistic population projections.
Further calibration and validation of these models using empirical estimates 
from cohort studies \citep{Zaba2007,Reniers2014} will be important in
this context.

\section{Age-Specific Fertility Rates for Probabilistic Population Projections}
\label{sect-fert}

\subsection{WPP 2012 Method of Projecting Age-Specific Fertility Rates}

The United Nations probabilistic population projections released in 2014 (\citealp{gerland&2014}) used a set of projected age-specific fertility rates for
each country obtained by combining probabilistic projections of the total fertility rate  with deterministic projections of age patterns of fertility as used in the 2012 revision of the World Population Prospects (\citealp{UN2014}).

For high-fertility and medium-fertility countries, future age patterns of fertility were obtained by interpolating linearly between a starting proportionate age pattern of fertility and a target model pattern.
The target model pattern was chosen from among 15 proportionate age patterns of 
fertility, with mean age at childbearing varying between 24 and 28.5 years.
The target pattern was held constant once the country reaches its lowest fertility level, or by 2045-2050 onward. 

For low fertility countries, a similar approach was used. This projected
future age-specific fertility patterns by assuming that they would reach
a target model pattern by 2025-2030.  This target was chosen from among
five target age patterns of fertility either for the market economies of 
Europe (with mean age of childbearing varying between 28 and 32 years) 
or for countries with economies in transition (with mean age of childbearing 
varying between 26 and 30 years). Once the model pattern was reached, it was assumed to remain constant until the end of the projection period. In some instances, a modified Lee-Carter approach (\citealp{Li&Gerland2009}) was used to extrapolate the most recent set of proportionate age-specific fertility rates using the rates of change from country-specific historical trends.

All the trajectories making up the probabilistic projection of fertility 
for a given country used the same age pattern of fertility. 
The choice of target pattern of fertility for a given country, 
from among the set of model patterns considered,
was driven by country-specific expert opinion about future trends 
and normative assumptions. 
No global or regional convergence in age patterns of fertility was imposed.

\begin{figure}
\begin{center}
\includegraphics[width=\textwidth]{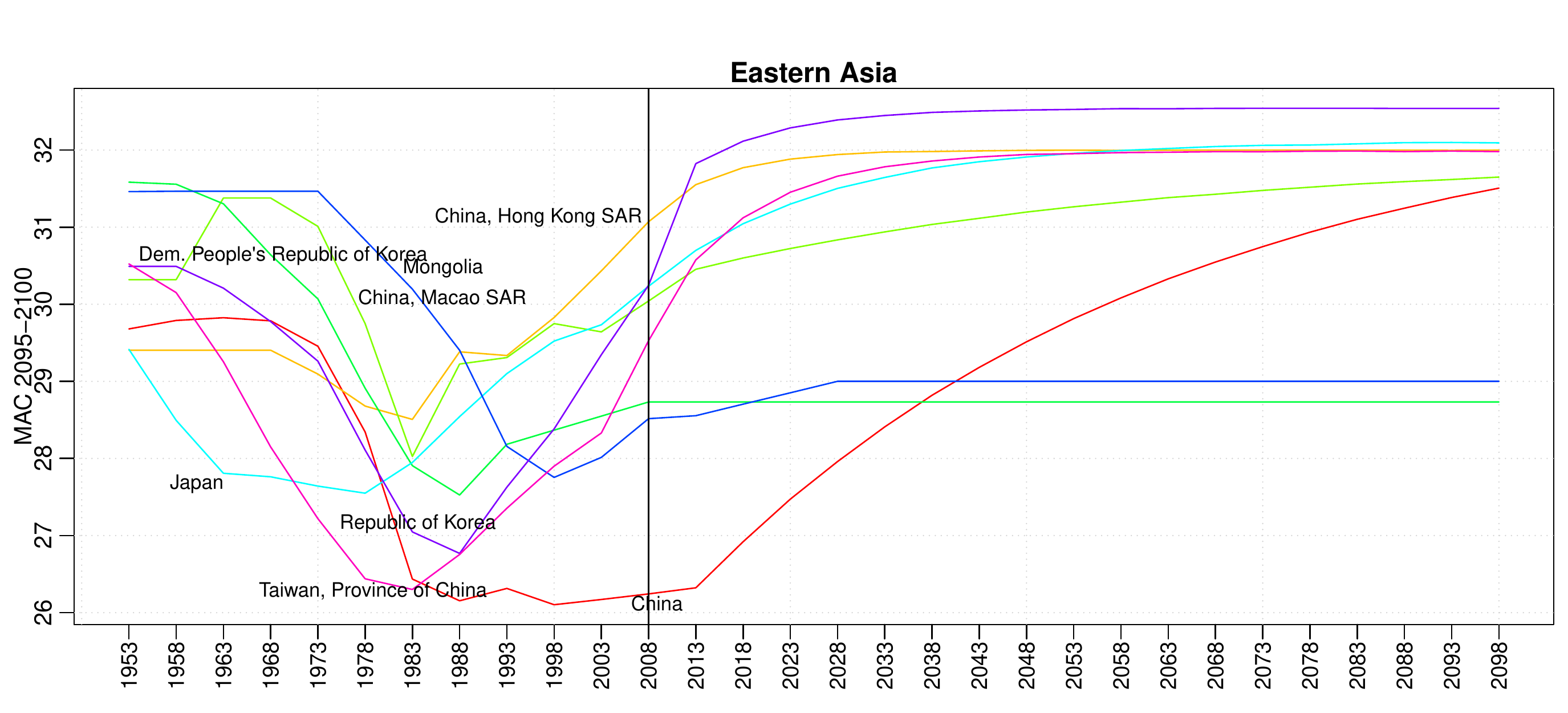}
\end{center}
\caption{\label{fig-MAC-WPP2012-EAsia} Example of projected Mean Age of Childbearing (MAC) for countries in Eastern Asia in WPP 2012.}
\end{figure}

Figure \ref{fig-MAC-WPP2012-EAsia} shows the results of the projections for the Mean Age of Childbearing (MAC) for countries in Eastern Asia from the 2012 Revision of the World Population Prospects. 

Overall, the method for projecting age-specific patterns of fertility in 
the 2012 Revision (as well as in previous revisions) has several limitations. 
First, no global or regional convergence has been imposed despite the overall
convergence in total fertility rates observed in the projection period up to
2100. Second, the time point when the target age-specific pattern is reached
is not related to the projected total fertility rates. Third, expert
assumptions on the target age pattern and method used for individual
countries introduce diversity in the age-specific trends are difficult
to explain (see Figure \ref{fig-MAC-WPP2012-EAsia} --- 
Mongolia and Democratic People's Republic of
Korea were done by Analyst 1, all other countries by Analyst 2).
Finally, since the analysts have used at least two different methods and 25 target age patterns of fertility, the documentation of the
decisions made for individual countries have been challenging. 

\subsection{Convergence Method for Projecting Age-Specific Fertility Rates}

We now propose a new method for projecting age-specific fertility rates,
to overcome some of the limitations of the existing method used in 
WPP 2012.
This builds on the approach adopted in sets of projections of married or in-union women of reproductive age (MWRA) (\citealp{UN2013MWRA}). Beginning from the most recent observation of the age pattern of fertility 
in the base period of projection, the projected age patterns of fertility are based on the past national trend combined with the trend towards the global model age pattern of fertility. The projection method is implemented on the proportionate age-specific fertility rates (PASFR) covering seven age groups from 15-19 to 45-49. The final projection of PASFRs for each age group is a weighted average of two preliminary projections: 
\begin{enumerate}
\item[(a)] the first preliminary projection, assuming that the PASFRs converge to the global model pattern, see Section~\ref{sec:global-pattern}; and
\item[(b)] the second preliminary projection, assuming the observed national trend in PASFRs continues into the indefinite future, see Section~\ref{sec:national-trend}.
\end{enumerate}
The method is applied to all the trajectories that make up of the 
probabilistic projection of total fertility rate for all countries, 
based on the historical data in the WPP 2012 Revision (\citealp{gerland&2014}).

We now define the preliminary projections that constitute our overall 
projection. We use different notation than in Section \ref{sect-mort},
so the same symbol may be used to denote different quantities in the
two sections.

\subsubsection{Trend towards the global model pattern}
\label{sec:global-pattern}

Let $t_r$ denote the base period of a projection and $t_{g}$ 
the year when the global model pattern is reached. 
For $t_r < t < t_{g}$, the proportion of the interval $[t_r, t_{g}]$
that has elapsed at time $t$ is 
\begin{equation*}
\tau_t = (t-t_r) / (t_{g} - t_r). 
\end{equation*}
Section~\ref{sec:tg} below gives details about how to estimate $t_g$.

Let  $p_r$ denote PASFR at the base period $t_r$, and let $p_{g}$ denote PASFR of the global model pattern.\footnote{In the bayesPop package the global model pattern is created as an average of most recent PASFRs for a set of countries (selected through a dummy variable in the {\tt vwBaseYear2012} dataset). For the purpose of the current analysis, the low fertility countries selected have already reached their Phase III and represent later childbearing patterns with mean age at childbearing close to or above 30 years in 2010-2015: Austria, the Czech Republic, Denmark, France, Germany, Japan, the Netherlands, Norway and the Republic of Korea. The specification of the countries used for the global model pattern can be changed in input file.} The projections at time $t$ of PASFR towards the global model pattern, denoted by $p_t^{I}$, is obtained by:
\begin{eqnarray} \text{logit} (p_t^{I}) &=& \text{logit} (p_r) + \tau_t \left[\text{logit} (p_{g}) - \text{logit}(p_r)\right] \end{eqnarray}
Then $p_t^{I}$ is renormalized so that it sums to unity for all time periods $t$.

\subsubsection{Continuing of observed national trend}
\label{sec:national-trend}
Let  $T$ denote the number of 5-year periods over which the model is fitted. Then  $t_{r-T}$ is the starting time period of the estimation and $p_{(r-T)}$ is PASFR at $t_{r-T}$.  $p_t^{II}$ is the projected PASFR at time $t$, assuming the past trend was to continue into the future under the following rule:
\begin{eqnarray} \text{logit}(p_t^{II})&=&\text{logit}(p_{r}) + \dfrac{t-t_{r}}{t_{r}-t_{r-T}} \left[\text{logit}(p_{r})-\text{logit}(p_{r-T})\right] \end{eqnarray}	
As above, $p_t^{II}$ should be scaled to sum to unity for all $t$.
Note that in our implementation we use $T=3$.

\subsubsection{Resulting projection}

Projected PASFR at time $t$, $p_t$, is calculated as:
\begin{eqnarray}\text{logit}(p_t) &=&  \tau_t \cdot {\text{logit}}(p_t^{I}) + (1 - \tau_t) \text{ logit}(p_t^{II} )\end{eqnarray}

Resulting  $p_t$ is renormalized  to sums to unity for all time periods $t$.

\subsubsection{Estimating the time period of reaching global pattern}
\label{sec:tg}

\begin{figure}
\begin{center}
\includegraphics[width=\textwidth]{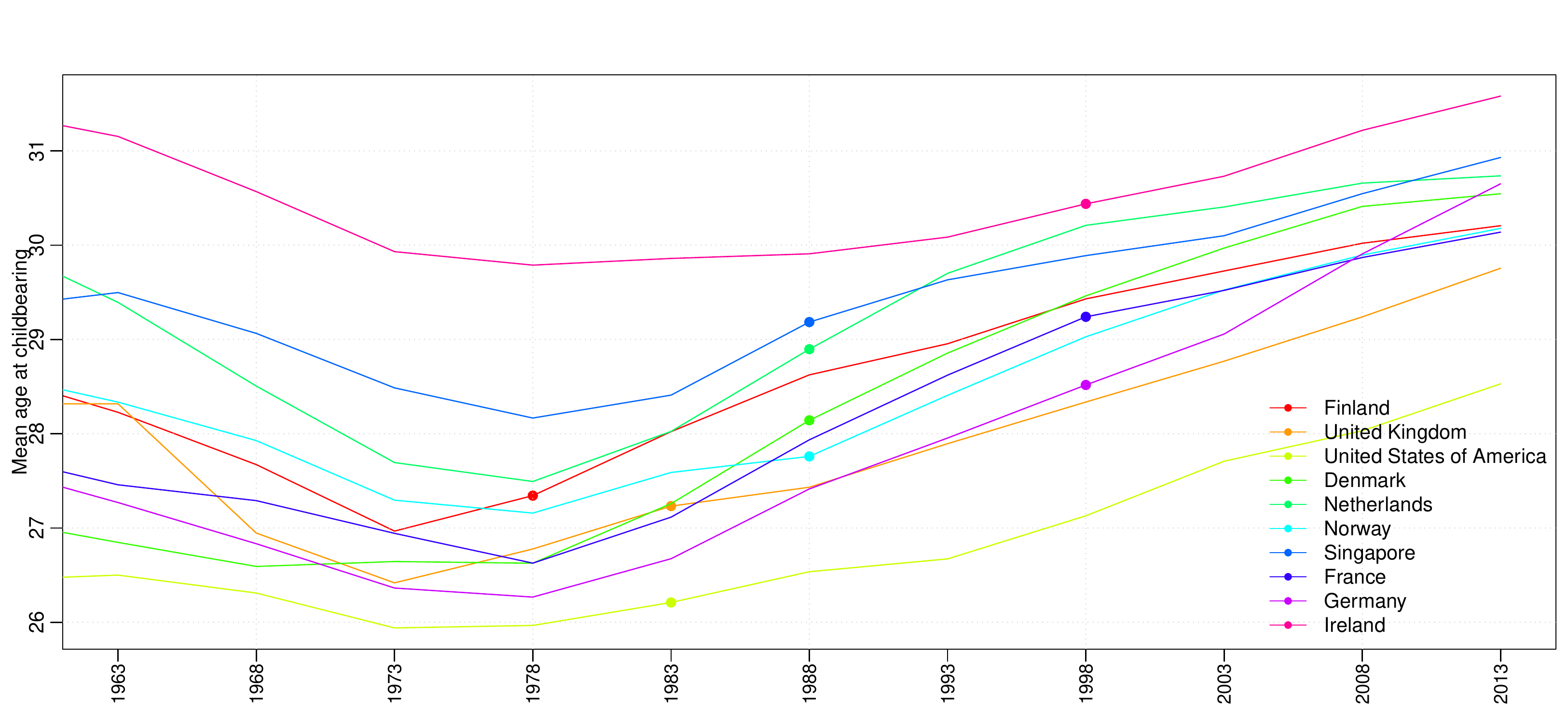}
\end{center}
\caption{\label{fig-MAC-PhaseIII} Trends in Mean Age at Childbearing in countries with the start of  Phase III of fertility decline before 2000. Dots mark the time period when the country entered Phase~III.}
\end{figure}

We assume that the transition from the most recent age pattern of fertility to the global model age pattern of fertility is dependent on the timing when the total fertility rate (TFR) enters Phase III, i.e. when the fertility transition is completed and the country reaches low fertility. For the countries in Phase III, a time series model to project TFR was used that assumed that in the long run fertility would approach and fluctuate around country-specific ultimate fertility levels based on a Bayesian hierarchical model (\citealp{Raftery&2014}). The time series model uses the empirical evidence from low-fertility countries that have experienced fertility increases from a sub-replacement level after a completed fertility transition. At the same time, based on the empirical evidence on the postponement of childbearing in low-fertility countries, profound shifts to later start of childbearing and an increase in the mean age of childbearing are still taking place several periods after the start of Phase III (see Figure~\ref{fig-MAC-PhaseIII}). The timing and speed of the postponement of childbearing in Phase III is country-specific and in this paper we implement the assumption that the transition to later childbearing pattern is completed when total fertility approaches country-specific ultimate fertility levels.

To be more specific, we assume that the time $t_g$ of a completion of the transition to a global model pattern corresponds to the time point $t_{u}$, when TFR reaches the ultimate fertility level of that country. In  probabilistic projections of TFR, we approximate the ultimate fertility level, denoted by $f_u$, by the median TFR in the last projection period $t_e$, e.g.  $t_e=$ 2095-2100, if TFR is in Phase III:
\begin{eqnarray}
\hat{f}_u & = & \text{median}_i\left[f_i(t_e)\right] \quad (i \text{ denotes trajectories})
\end{eqnarray}
Then  for each TFR trajectory, $t_u$ is the earliest time period, at which the TFR  is larger or equal to $\hat{f}_u$:
\begin{eqnarray}
t_u &=& \min\{t : f(t)  \geqslant  \hat{f}_u \; \text{ and } \; t > t_{P3}\} 
\end{eqnarray}
where $t_{P3}$ denotes the start of Phase III. For the estimation of $t_g$, we will now distinguish two cases, depending if $t_{P3}$ is smaller or larger than the end period $t_e$.

\paragraph{Case 1: $t_{P3} < t_e$}~\\[3mm]
In this case, $t_{P3}$ is either observed ($t_{P3} \leqslant t_r$) or projected within the projecting period ($t_r < t_{P3} < t_e$). In both cases, if $t_u$ exists,
\begin{eqnarray} t_{g} &=& \max (t_{u}, t_r + 10)\,. \end{eqnarray}
This includes a situation where $f(t ) \geqslant  \hat{f}_u$ for $t \leqslant  t_r$. In such a case, the global pattern is reached quickly, namely in two 5-year periods.

If $f(t) < \hat{f}_u$  for all $t_r \leqslant t \leqslant t_e$, then $t_u$ does not exist. In such a case, $t_g$ is set to the end of the projection period, but at least five 5-year periods after $t_{P3}$:
\begin{eqnarray} \label{eq:case1} t_{g} &=& \max(t_e,  t_{P3} + 25) \end{eqnarray}


\paragraph{Case 2: $t_e \leqslant t_{P3}$}~\\[3mm]
In this case, $t_{P3}$ is unknown, i.e. the TFR trajectory has not reached Phase III at $t_e$. Thus, we will make an estimate of $t_{P3}$, denoted by $\hat{t}_{P3}$, and then simply apply 
\begin{eqnarray} \label{eq:case2} t_{g} &=& \hat{t}_{P3} + 25 \,. \end{eqnarray}
If the TFR at $t_e$ is low, namely $f(t_e) \leqslant 1.8$, we assume that $\hat{t}_{P3}=t_e$.
Otherwise, we approximate $t_{P3}$ by a linear extrapolation of TFR from the last four time periods and determine when such line reaches 1.8, with an upper limit of $\hat{t}_{P3} = t_e + 50$.


\subsubsection{Exception for late childbearing pattern}
Since trajectories for some countries have already observed or -- as projected by the algorithm described above -- will in near future reach higher MAC than the MAC associated with the global model pattern, we assume that for a given country's trajectory once the maximum MAC is reached in the convergence period the associated PASFR pattern is kept constant for the remaining projection periods. This assumption enables to keep trajectory-specific patterns of late childbearing for trajectories after the Phase III, thus already with low total fertility  (see Figure~\ref{fig-PASFR} for example of the Czech Republic). Note that this rule is applied only in Case 1 above.

\subsection{Results of the convergence method applied to probabilistic projections}
For the 2012 Revision, age-specific fertility estimates are based on empirical data for all countries of the world for the period up to 2010 (or up to 2010-2015 for 37 countries with empirical data up to 2011 or 2012; \citealp{gerland&2014}). Using the probabilistic projections of TFR, each TFR trajectory has a specific start of Phase III and therefore the timing of convergence to the global model pattern is trajectory-specific. This yields a set of trajectories of PASFR (although not probabilistic) which in turn, when combined with the probabilistic TFR, yield probabilistic projection of age-specific fertility rates.

\begin{figure}[h]
	\begin{center}
		\includegraphics[width=\textwidth]{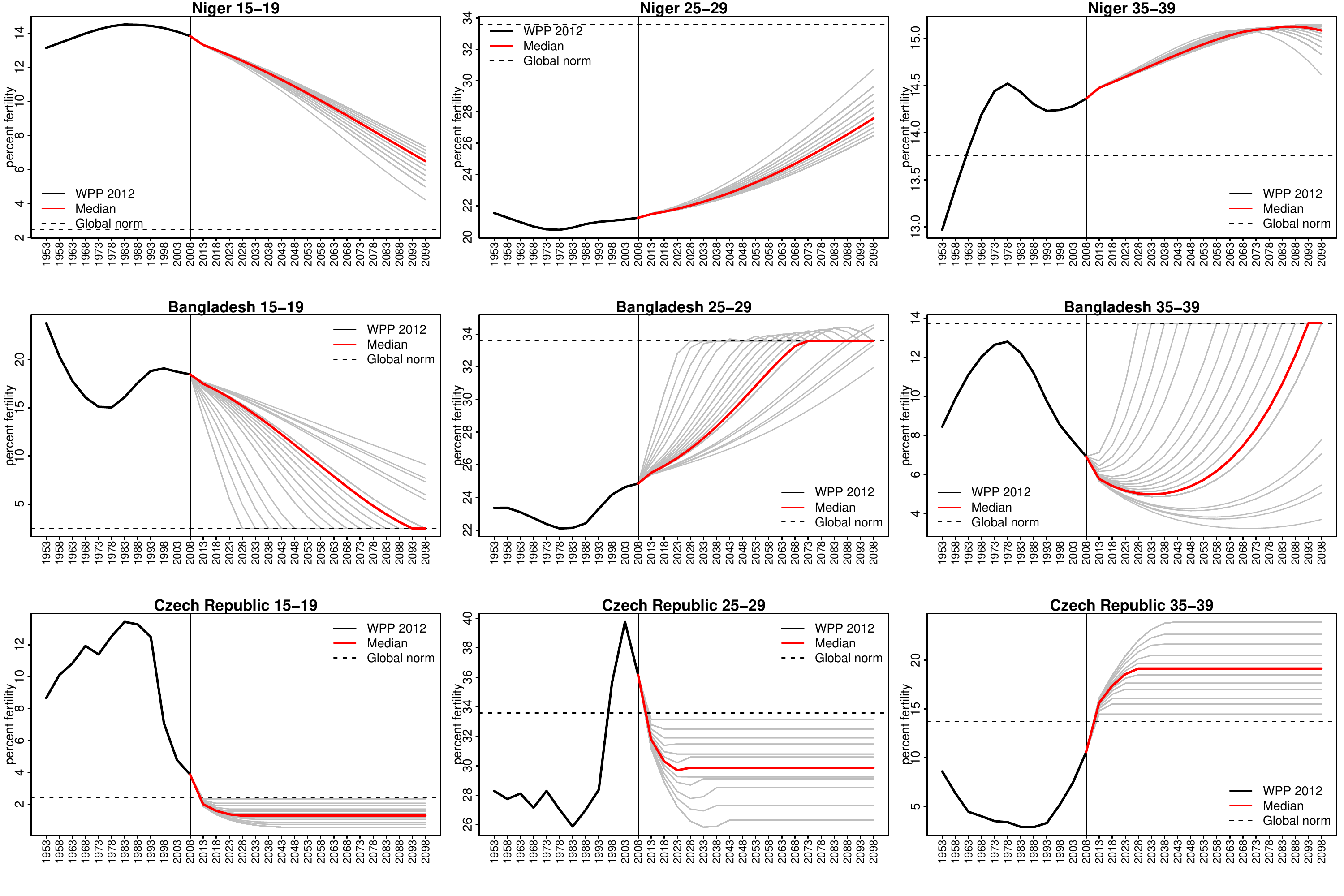}
	\end{center}
	\caption[Proportionate age-specific fertility rates (PASFR) by time for age groups 15-19, 25-29 and 35-39 in Niger, Bangladesh and the Czech Republic]
{\label{fig-PASFR} Proportionate age-specific fertility rates (PASFR) by time for age groups 15-19, 25-29 and 35-39 in Niger, Bangladesh and the Czech Republic. Projected median of PASFR (red line) approaches global model pattern of PASFR (black dashed line). 
The solid grey lines are trajectories that correspond
to different starting periods for Phase III; they do not represent
random samples from a predictive probability distribution.
}
\end{figure}

\begin{figure}[h]
	\begin{center}
		\includegraphics[width=\textwidth]{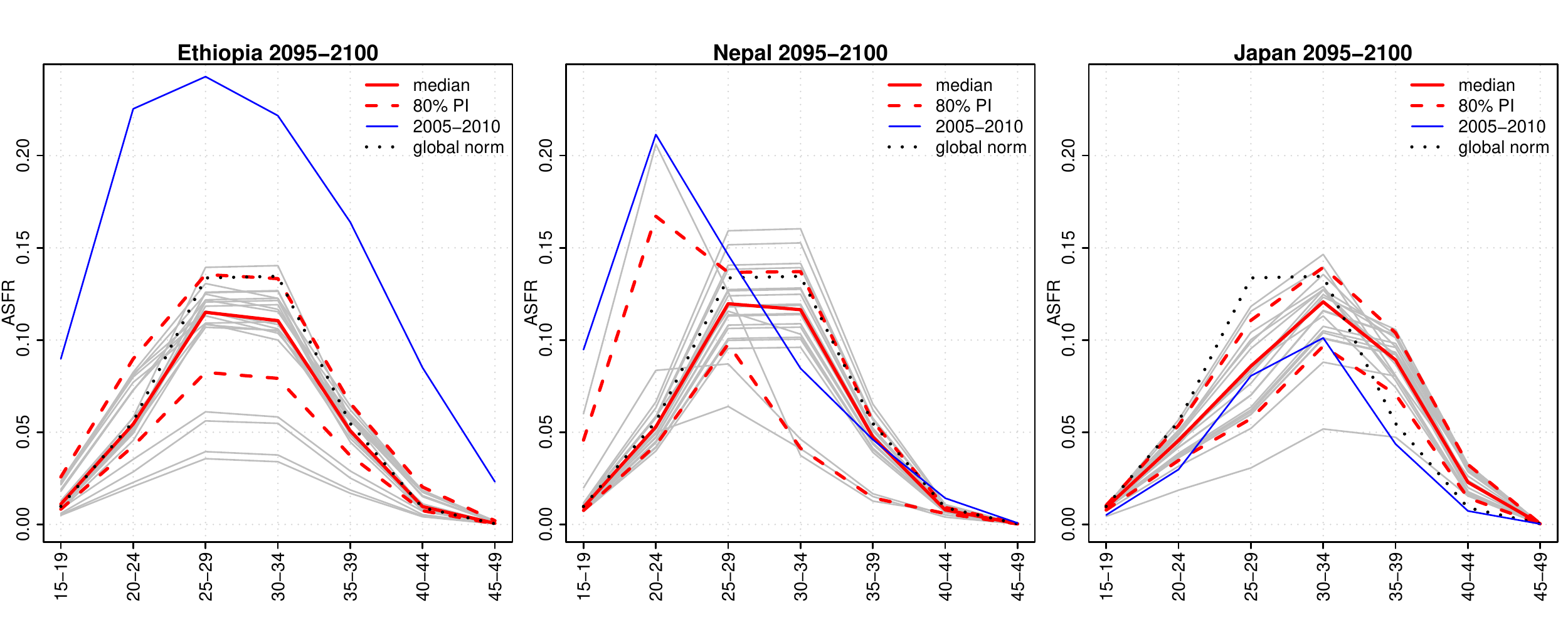}
	\end{center}
	\caption[Probabilistic projection of age-specific fertility rates for Ethiopia, Nepal and Japan in the time period 2095-2100]
{\label{fig-ASFR2095-2100} Probabilistic projection of age-specific fertility rates for Ethiopia (left panel), Nepal (middle panel) and Japan (right panel) in the time period 2095-2100. The marginal distribution for age-specific fertility rates (red lines) where the dashed lines mark the 80\% probability intervals and the solid grey lines are randomly sampled trajectories are compared to age-specific fertility rates in the time period 2005-2010 (blue line) and to the global model pattern applied to median projection of total fertility for the world in 2095-2100 (black dashed line).}
\end{figure}

Figure \ref{fig-PASFR} shows an example of the results for PASFR in Niger, Bangladesh and Czech Republic for selected age groups over time. Figure \ref{fig-ASFR2095-2100} shows an example of the probabilistic results of age-specific fertility rates for Ethiopia, Nepal and Japan at the end of projection period in 2095-2100.

\begin{figure}[h]
\begin{center}
\includegraphics[width=\textwidth]{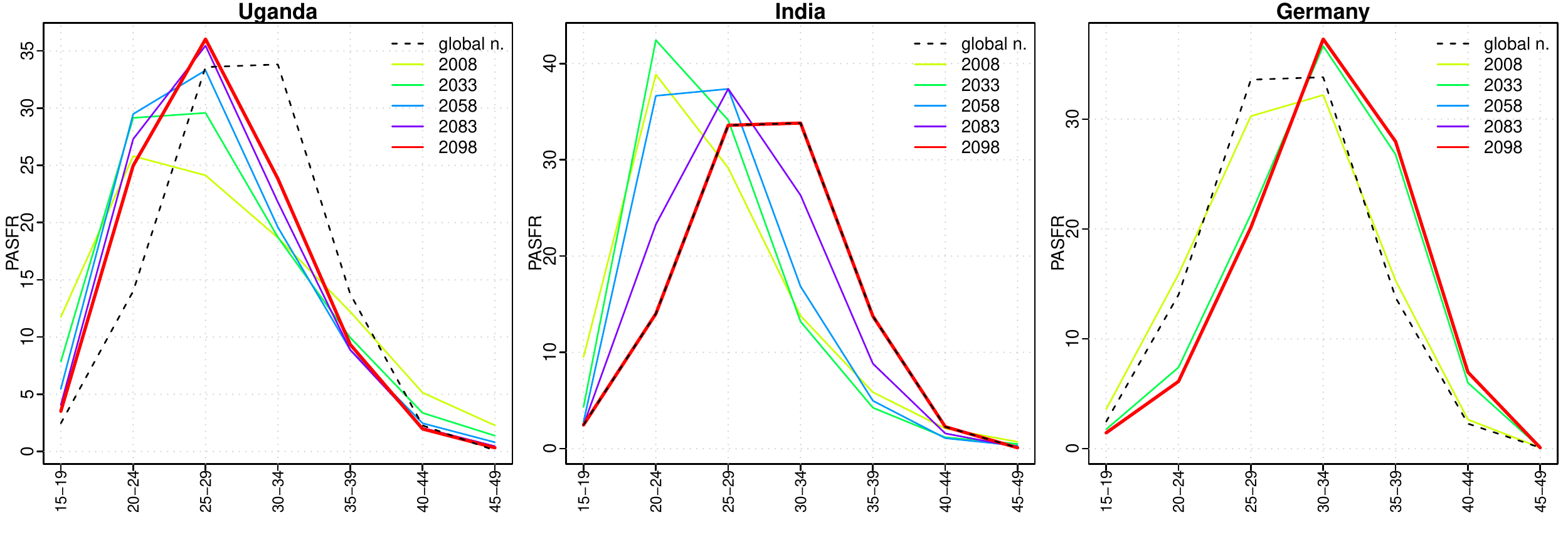}
\end{center}
\caption{\label{fig-PASFR-over-time} PASFR by age over time for selected countries.}
\end{figure}

Figure~\ref{fig-PASFR-over-time}  shows the development of PASFR for Uganda, India and Germany over time from 2005-2010 to 2095-2100. Here, the methodology was applied to the deterministic projection of TFR from WPP 2012.

\begin{figure}[h]
\begin{center}
\includegraphics[width=\textwidth]{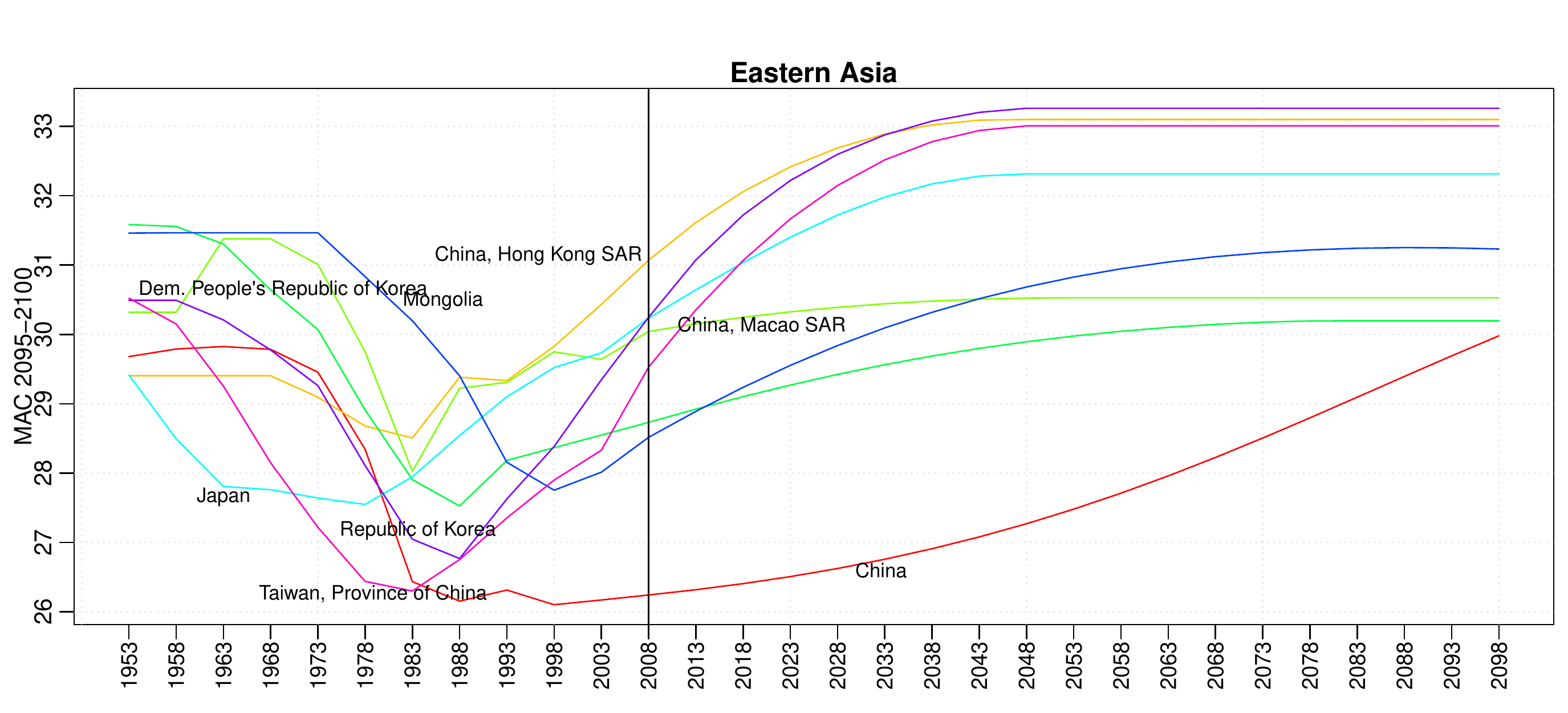}
\end{center}
\caption{\label{fig-MAC-NEW-EAsia} Example of projected MAC for countries in Eastern Asia after applying the proposed methodology.}
\end{figure}

In Figure~\ref{fig-MAC-WPP2012-EAsia} we showed projections of MAC from WPP 2012. This can be compared to 
Figure~\ref{fig-MAC-NEW-EAsia} where the same measure is shown after  applying the new methodology to the TFR of WPP 2012.

Overall, the new method we propose improves on the current methodology 
in several ways.
First, in the very long term (after 2100) the age patterns of fertility are
converging to one global pattern, while retaining specific late childbearing
patterns for several countries that reach such patterns in the current period 
or in the near future.
Second, the projections of the age pattern of fertility are
now linked to projections of the total fertility rate. 
Finally, for each
probabilistic trajectory, the time when the target age pattern is reached
depends on the trajectory-specific total fertility rate.

\section{Discussion}
\label{sect-discussion}

We have described the methods used for converting projected
life expectancies at birth and total fertility rates 
to age-specific mortality and fertility rates in the UN's 2014 
probabilistic population projections. We have identified some limitations
of these methods and have proposed several improvements to overcome them.
These include a new coherent Kannisto method to avoid crossovers
in mortality rates between the sexes at very high ages.
They also include the application of a coherent Lee-Carter method, methods for
avoiding jump-off bias, and a rotated Lee-Carter method to reflect
the fact that at high life expectancies, mortality rates tend to decline
faster at higher than at lower ages.

It should be noted that the 2014 PPP takes account of uncertainty
about the overall level of fertility as measured by the TFR,
and also about the overall level of mortality as measured by $e_0$.
Conditional on TFR and $e_0$, however, the projected vital rates are
deterministic. There is thus a missing component of uncertainty,
and it would be desirable to extend the methods used to take account of this,
particularly of uncertainty about the future mean age at childbearing
\citep{Ediev2013}.

\section{Acknowledgements}
This research was supported by NIH grants R01 HD054511 and R01 HD070936. 
The views expressed in this article are those of the authors and do not 
necessarily reflect those of NIH or the United Nations.

\clearpage

\bibliography{SevcikovaArxiv}
\bibliographystyle{newapa}
\end{document}